\documentclass[12pt]{JHEP3}

\let\ifpdf\relax
\usepackage{ifpdf}
\usepackage{graphics}
\usepackage{graphicx}
\usepackage{amssymb}
\usepackage{amsmath}
\usepackage{slashed}
\usepackage{cite}
\usepackage{epsfig}
\usepackage{datetime}

\newcommand{\be}{\begin{equation}}
\newcommand{\ee}{\end{equation}}
\newcommand{\bal}{\begin{align}}
\newcommand{\eal}{\end{align}}
\newcommand{\bea}{\begin{eqnarray}}
\newcommand{\eea}{\end{eqnarray}}

\def\O{\mathcal{O}}

\def\Tr{{\rm Tr}}

\def\tr{{\rm tr}}

\def\O{\mathcal{O}}

\title{Quantum Corrections to Holographic Mutual Information}

\preprint{
 BRX-TH-6299}
\author{Cesar A. Ag\'on${}^{*,a}$  and Thomas Faulkner${}^{\dagger,b}$\\
{\small ${}^*$ Martin Fisher School of Physics, Brandeis University, \\ \ \ \ \ \ Waltham, MA 02453, USA\\
\small ${}^{\dagger}$ University of Illinois, Urbana-Champaign, \\ \ \ \ \ \ Urbana, IL 61801-3080, USA\\
}
${}^a$\email{caagon87@brandeis.edu}\\
\ ${}^b$\email{tomf@illinois.edu}
}
\abstract{ \\ We compute the leading contribution to the mutual information (MI) of two disjoint spheres in the large distance regime for arbitrary conformal field theories (CFT) in any dimension.  This is achieved by refining the operator product expansion method introduced by Cardy \cite{Cardy:2013nua}. 
For CFTs with holographic duals the leading contribution to the MI at long distances
comes from bulk quantum corrections to the Ryu-Takayanagi area formula. According to the FLM proposal\cite{Faulkner:2013ana} this
equals the bulk MI between the two disjoint regions spanned by the boundary spheres and their corresponding minimal area surfaces. We compute this quantum correction 
and provide in this way a non-trivial check of the FLM proposal. 
}

\begin{document}

\section{Introduction}

\begin{figure}[h!]
\centering
\includegraphics[trim={0 0 0 .7cm}, scale=.5, clip]{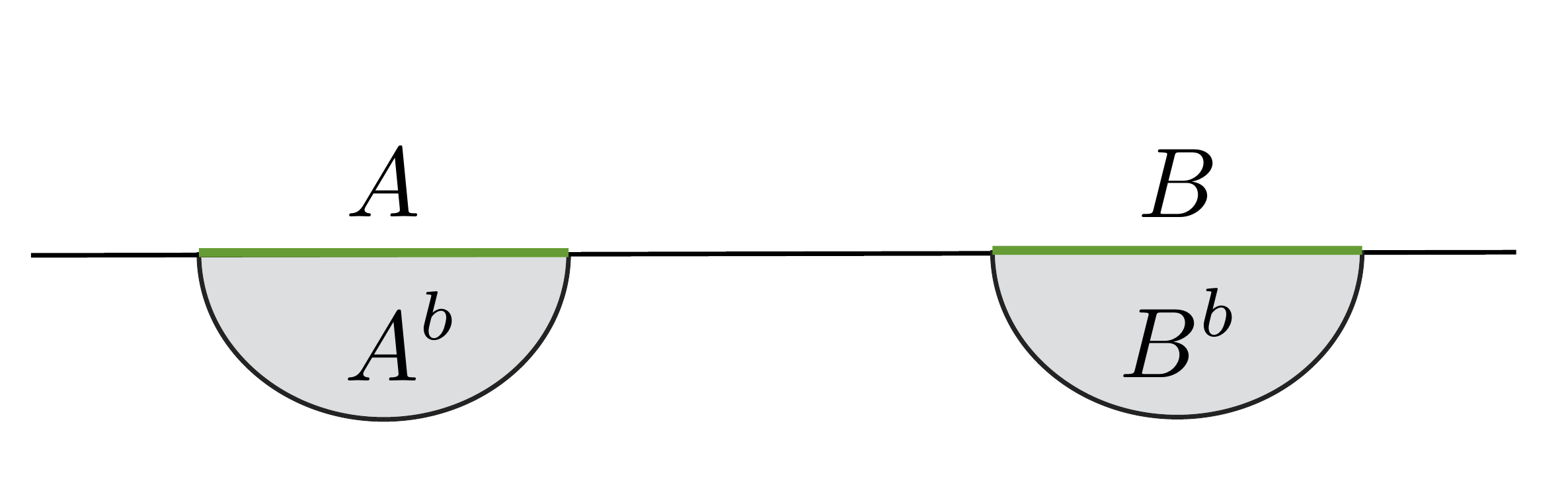}
\caption{This paper studies the MI between the boundary regions $A,B$. The FLM proposal predicts the non-zero contribution to the MI for $r > r_c$ is given by the bulk MI between the hemispherical regions $A_b$ and $B_b$ contained within the Ryu-Takayanagi minimal surfaces. \label{flm} }
\end{figure}

It is well known that the mutual information for disjoint and compact regions in holographic theories undergoes a sharp transition when the separation distance is larger than some characteristic scale $r_c$ \cite{Headrick:2010zt}. The usual Ryu-Takayanagi formula for the entanglement entropy gives a zero contribution to the mutual information for $r > r_c$ and therefore we expect the leading non-zero answer to be determined by quantum fluctuations in the dual space-time. As proposed by Faulkner, Lewkowycz and Maldacena (FLM) \cite{Faulkner:2013ana} such contributions at leading order are given by the mutual information between the bulk regions depicted in Figure~\ref{flm}.

\vspace{-.4cm}

While FLM gives us an in principle prescription for calculating quantum corrections to entanglement entropy actually carrying out such computations is technically challenging.\footnote{
Some previous applications of FLM include \cite{Swingle:2013rda,Leichenauer:2015nxa,Swingle:2014uza,Miyagawa:2015sql}. We should also mention that an important generalization of FLM to higher orders in the bulk quantum expansion was given in \cite{Engelhardt:2014gca}.}
In three bulk dimensions alternative methods to compute quantum corrections to MI are available  \cite{Barrella:2013wja}. This approach is based on computing the one-loop determinant of the bulk partition function in the geometries constructed in \cite{Faulkner:2013yia}, and
has been extensively and successfully applied in a large variety of situations \cite{Chen:2015uia,Chen:2015kua,Chen:2014hta,Chen:2014unl,Beccaria:2014lqa,Chen:2014kja,Chen:2013kpa,Chen:2013dxa,Perlmutter:2013paa} finding agreement with independent CFT calculations.
It should be emphasized that the approach of \cite{Barrella:2013wja} is quite different to FLM. It is expected that the two approaches should agree for 2d CFTs, however this has never been explicitly demonstrated.  Additionally, the one loop determinant methods are essentially out of reach in higher dimensions.
It is thus of fundamental importance to develop techniques that allow us to carry out computations with the FLM proposal, and to check the results against boundary CFT calculations where available.

An obvious difficulty this program faces is the lack of CFT results for entanglement entropy/mutual information in  CFT's  that one could use to compare with their corresponding holographic predictions. 
We  plan to remedy this situation and the first result we would like to present is the leading correction to MI in the limit of large distances $r$ between the two spherical shaped regions in any CFT\footnote{With the one condition that the lowest operator dimension in the CFT is a scalar. Presumably a very similar
result holds for spinors, vectors and the stress tensor as suggested from the two dimensional case \cite{Perlmutter:2013paa}, however we leave this for future work. }:
\be
I =  \mathcal{N}_\Delta \frac{\sqrt{\pi}\Gamma(2\Delta +1)}{4\Gamma(2\Delta +\frac 32)}\frac{(R_AR_B)^{2\Delta}}{r^{4\Delta}} + \ldots
\label{eq:result}
\ee
where $\Delta$ is the scaling dimension of the lowest dimension scalar operator, $R_A$ and $R_B$
are the radius of the two spheres and $\mathcal{N}_\Delta$ is the number/degeneracy of 
such real scalar operators. 
We use the framework setup by Cardy for calculating MI in higher dimensions \cite{Cardy:2013nua}.
While the scale dependent part of \eqref{eq:result} was established in \cite{Cardy:2013nua}
and in the earlier numerical work of \cite{Shiba:2012np}, the exact pre-factor was left unknown, except for some results in free theories. Surprisingly the pre-factor we find is the same as in $2d$ CFTs  \cite{Calabrese:2010he}. 

The methods of  \cite{Cardy:2013nua} have been applied in a variety of situations to compute R\'enyi mutual information and mutual information for free scalars and fermions at zero and finite temperature \cite{Schnitzer:2014zva,Herzog:2014fra,Herzog:2014tfa,Agon:2015twa,Herzog:2015cxa}. 
The main tool is an operator product expansion argument used to express the R\'enyi mutual information in terms of  multi-correlators of the different fields in the replicated geometry \cite{Headrick:2010zt,Calabrese:2010he,Cardy:2013nua}. The new ingredient we add to this discussion is a method to find the analytic continuation in the replica parameter $n$ of sums of the OPE coefficients that works for any CFT. 
This continuation is inspired by recent results for computing perturbative corrections to EE \cite{Faulkner:2014jva, Faulkner:2015csl}.

Turning to the bulk FLM computation we find that the framework of \cite{Cardy:2013nua} can also be used to compute the leading term of the bulk mutual information for disjoint hemispheres in arbitrary space-time dimensions. We consider only the contribution coming from a free scalar field living on the $AdS_{d+1}$ background, where the mass of the scalar is related to the conformal dimension of the dual operator in the usual way. Interestingly, the curved geometry does not represent any obstacle in carrying out this calculation and therefore opens the door to a larger exploration of mutual information in curved geometries. These two independent results agree perfectly and therefore provide important evidence for the validity of FLM.

The organization of this paper goes as follows: in section \ref{2}, we present a brief overview of the general framework for computing mutual information of disjoint regions in a large distance expansion, following closely the presentation of \cite{Cardy:2013nua}. In section \ref{3} we present a detailed calculation of the coefficient of the leading term in the mutual information for disjoint spheres in arbitrary CFT and for any dimension. The equivalent dual bulk computation, that is the mutual information between hemispheres in the AdS background, is presented in the section \ref{4}.

\section{Mutual Information expansion \label{2}}
In this section we briefly review the general framework to compute the mutual information between disjoint regions in quantum field theories following closely \cite{Cardy:2013nua}. We take the QFT to live on a (potentially) curved $d$ dimensional manifold $\mathcal{M}$.

The R\' enyi entropy for a given region $X$ is given by
 \bea
S^{(n)}_X=\frac{1}{1-n}\log \Tr_{{\cal H}_X}\rho^n_X
\eea
where ${\cal H}_X$ is the Hilbert space associated to the region $X$ and $\rho_X$ is the reduced density matrix describing the degrees of freedom living in $X$. When $n$ is an integer this quantity can be expressed in terms of a path-integral on a conifold defined by taking $n$ copies of the QFT 
on $\mathcal{M}^{\otimes n}$ and sewing them together along the region $X$. The R\'enyi entropies are given by
 \bea
S^{(n)}_X=\frac{1}{1-n}\log\left(\frac{Z({\cal C}^{(n)}_X)}{Z^n}\right)\,,
\eea
where $Z$ is the partition function on the original space, and $Z({\cal C}^{(n)}_X)$ is the partition function on the conifold.
We are interested in the mutual information $I(A,B)$ associated to two disjoint regions $A$ and $B$, which can be defined as
\bea
\label{lim}
I(A,B)\equiv \lim_{n\to 1} I^{(n)}(A,B)
\eea
where $I^{(n)}(A,B)$ is the  R\'enyi mutual information  given by:
\bea
I^{(n)}(A,B)\equiv S_A^{(n)}+S_B^{(n)}-S_{A\cup B}^{(n)}
\eea 
or in terms of path integrals
\bea
\label{renyiinfo}
I^{(n)}(A,B)=\frac{1}{1-n}\log\left(\frac{Z({\cal C}^{(n)}_{A\cup B})Z^{n}}{Z({\cal C}^{(n)}_A)Z({\cal C}^{(n)}_B)}\right)\,.
\eea
Part of the difficulty of this calculation is coming up with an analytic continuation in $n$ away from the integers in order to properly take the limit \eqref{lim}. We will address this issue shortly.
Further simplification occurs in the limit where the distance between objects $r$ is much larger than the individual sizes $R_A,R_B$. In that situation we can think of the sewing operation in the region $B$ as seen from the point of view of $A$ as a semi-local operation that couples the n QFTs. That is 
\bea
\frac{Z({\cal C}^{(n)}_{A\cup B})}{Z^{n}}=\langle  \Sigma_A^{(n)} \Sigma_B^{(n)} \rangle_{\mathcal{M}^n}
\eea 
where we make the replacement:
\bea
\Sigma_{A}^{(n)}=\frac{Z({\cal C}_A^{(n)})}{Z^n}\sum_{\{k_j\}}C_{\{k_j\}}^{A} \prod_{j=0}^{n-1} \Phi^{(j)}_{k_j}(r_A)
\eea
and $\Phi^{(j)}_{k_j}(r_A)$ is a complete set of operators in the $j$th copy of the QFT located
at a conveniently chosen point $r_A$ in region $A$.\footnote{For example an operator that is displaced from $r_A$ arise as an infinite sum over derivatives of this operator located at $r_A$.} 
By making appropriate subtractions we can take these operators to have vanishing one point functions on $\mathcal{M}$.
An arbitrary product of operators far from $A$ in the conifold geometry of region $A$ 
is therefore given by
\bea
\label{correlators}
\langle \prod_{j'=0}^{n-1} \Phi^{(j')}_{k'_{j'}}(r)\rangle_{{\cal C}^{(n)}_A}&=&\langle \prod_{j'=0}^{n-1}\Phi^{(j')}_{k'_{j'}}(r) \sum_{\{k_j\}}C_{\{k_j\}}^{A} \prod_{j=0}^{n-1} \Phi^{(j)}_{k_j}(r_A)\rangle_{\mathcal{M}^{\otimes n}}\nonumber \\
&=&\sum_{\{k_{j}\}}C_{\{k_{j}\}}^{A}\prod_{j} \langle \Phi_{k'_{j}}(r)\Phi_{k_{j}}(r_A)\rangle_{\cal M}
\eea
As pointed out by Cardy this equation is true for any arbitrary QFT, however for a CFT in flat space we can use scalar operators with the following normalization
\bea
\label{normalization}
\langle \Phi_{k'}(r) \Phi_k (r_A)\rangle_{\mathbb{R}^d} =\frac{\delta_{k k'}}{|r-r_A|^{2x_k}}\,.
\eea
Plugging (\ref{normalization}) into (\ref{correlators}) allows us to extract the coefficients of interest
\bea
\label{coeffs}
C_{\{k_{j}\}}^{A}=\lim_{r \to \infty}|r|^{2\sum_j x_{k_j}} \langle \prod_{j'} \Phi^{(j)}_{k_{j}}(r)\rangle_{{\cal C}^{(n)}_A}\,
\eea
and with them we can write a formal expression for the ratio of partition functions required for the evaluation of the R\'enyi mutual information \footnote{ See \cite{Cardy:2013nua} for further details.}
\bea
\label{insidelog0}
\frac{Z({\cal C}^{(n)}_{A\cup B})Z^n}{Z({\cal C}^{(n)}_{A})Z({\cal C}^{(n)}_{B})}=\sum_{ \{k_j\}}{ C^A_{\{k_j \}}}{ C^B_{\{ k_j \}}} r^{-2\sum_j x_{k_j}}\,.
\eea
Assuming a $\mathbf{Z}_2$ symmetry\footnote{This assumption can be relaxed and does not effect the answer as we take $n \rightarrow 1$ later on. For simplicity of the presentation we do not treat the non-symmetric case explicitly.} for the lowest dimension operator in the CFT, say $\O$,  (  $\mathbf{Z}_2$ acts as $\O(r)\to -\O(r)$) such that the symmetry is not spontaneously broken in the replica manifold, then $\langle\O^{(j)}(r) \rangle_{{\cal C}^A_n}=0$. Thus the first contribution to (\ref{insidelog0}) comes
from replacing each of the $\Sigma^{(n)}_{A,B}$ with two operator insertions of $\O$ on the different replicas $j,j'=0, \ldots n-1$ and is therefore given by 
\bea
\label{insidelog1}
\frac{Z({\cal C}^{(n)}_{A\cup B})Z^n}{Z({\cal C}^{(n)}_{A})Z({\cal C}^{(n)}_{B})}=1+\frac{1}{2}\sum_{j \neq j'}{ C^A_{jj'}}{ C^B_{jj'}} r^{-4 \Delta} + \ldots\,,
\eea
where the factor of $1/2$ appears to account for the double counting in the sum, and the number 1 comes from the contribution to (\ref{coeffs}) where all $\Phi_k(r)=\mathbf{1}$. 

Using \eqref{coeffs} the coefficients $C_{jj'}$ are simply
\bea
\label{coeffs2}
C_{jj'}^{A}=\lim_{r \to \infty } |r|^{4 \Delta} \langle \O^{(j)}(r) \O^{(j')}(r) \rangle_{{\cal C}^{(n)}_A}\,.
\eea
where $\Delta$ is the lowest scaling dimension of the CFT operators and then (\ref{insidelog1}) just depends on two point functions in the conifold manifold.
The first non-trivial contribution to the mutual information is
\bea
\label{want}
I(A,B)&=&\lim_{n\to 1}\frac{1}{1-n}\left(\frac n2 \sum_{j=1}^{n-1} { C^A_{0j}}{ C^B_{0j}} \right)r^{-4 \Delta}
+ \ldots
\eea
where we have used the cyclicity of the replica manifold to reduced the double sum to a single one.

\section{CFT in Euclidean space \label{3}}

In this section we evaluate the leading term in the mutual information (\ref{want}), when $A \cup B$ is a system of two largely separated $d-1$ spheres. In our setup we will consider an arbitrary CFT in flat euclidean geometry with $d$ dimensions. Our starting point uses the conformal transformation introduced in \cite{Casini:2011kv}, to map two point correlation functions 
in the conifold associated to a single entangling sphere to finite temperature two point functions in Hyperbolic space. As shown in \cite{Faulkner:2014jva}, sums over the replica manifold of thermal Green functions can be computed by using their analyticity properties as well as its exact form for ($n=1$), which is the one that can be conformally mapped to a two point function in flat euclidean space. We use a method similar
to that of \cite{Faulkner:2014jva}, to evaluate the limiting sum in (\ref{want}). 

Thinking in terms of embedding coordinates (we use the notation and conventions given in \cite{Faulkner:2014jva})  the map between  $\mathbb{R}^{d}$ and $\mathbb{S}^1\times \mathbb{H}^{d-1}$ can be stated as follow
\bea
\label{CT}
P|_H=\Omega P |_E
\eea
where $P$ is a point lying in the upper projective light cone of $\mathbb{R}^{1,d+1}$ with $P|_{E}$, $P|_{H}$ corresponding to its representation in terms of Euclidean and Hyperbolic coordinates respectively, and $\Omega=R^{-1}(Y^I +\cos\tau)$ is the conformal factor that relates them both. 
More specifically 
\bea
P |_E=\left(\frac{R^2+x^2}{2R},\frac{R^2-x^2}{2R},x^\mu \right), \qquad x^\mu \in \mathbb{R}^{d}
\eea
and 
\bea
P |_H=\left(Y^I, \cos \tau, \sin \tau, Y^m\right), \qquad (Y^I,Y^m) \in \mathbb{H}^{d-1}
\eea
where $\mathbb{H}^{d-1}$ is defined as the locus $-(Y^I)^2+Y^mY^m=-1$ and $Y^I>0$ with $m=1,\cdots d-1$, and $\tau$ is the coordinate on the $\mathbb{S}^1$ where $\tau \equiv \tau + 2\pi$.
While this conformal mapping is appropriate for a CFT living on flat space, it can be extended to the
conifold geometry ${ \cal C}^{(n)}_A$ simply by making the $\tau$ circle larger such that $\tau \equiv \tau+2\pi n$.

At the level of the two point functions in the conifold geometry the statement is
\bea
\label{CTcorrelators}
\langle {\cal O}(x_a) {\cal O}(x_b) \rangle_{{ \cal C}^{(n)}_A}=\Omega^{\Delta}(a)\Omega^{\Delta}(b) G_n(\tau_b -\tau_a ; Y_a\cdot Y_b)
\eea
where $G_n$ is the thermal Green function in hyperbolic space associated to the operator $\cal O$ at temperature $T=1/2\pi n$. That is
\bea
\label{TGF}
G_n(\tau_b -\tau_a ; Y_a\cdot Y_b)=\tr \left(e^{-2\pi n H} {\cal T} {\cal O}(i\tau_a,Y_a) {\cal O}(i\tau_b,Y_b)\right)
\eea
where $\cal T$ is for the euclidean time ordering operation.
We want to use this relation to evaluate the leading contribution to (\ref{want}).
Note that while the conformal factors are only well defined if $n$ is an integer, the
thermal green's function is defined for any $n$ and this will be the key feature that allows us
to analytically continue the sums in \eqref{want}. Further, while $G_n$ is an unknown theory dependent function, all we will need in order to compute the MI is the thermal greens function at $n=1$ which is known.

The explicit expression for the thermal Green function at $T=1/2\pi$ is,
\bea
G_1(\tau;Y_a\cdot Y_b)=\frac{c_\Delta}{(-2Y_a\cdot Y_b-2\cos(\tau))^\Delta}\,,
\eea
where it is easy to show that $c_\Delta = 1$ due to the normalization given in (\ref{normalization}). 
The point $r$ far away from region $A$ on the $j$'th replica maps under \eqref{CT} to:
\be
\left(\tau = \pi(2 j-1), Y_I \rightarrow 1 \right)  \quad {\rm with} \quad |r| \sim  2^{1/2} R (Y_I -1)^{-1/2} 
\ee

This allows us to write the OPE coefficients (\ref{coeffs2}) in terms of the thermal Green functions
\bea
C_{jj'}&=&\lim_{Y^I\to1}(R)^{4\Delta} 2^{2\Delta} (Y_I-1)^{-2\Delta}\Omega^{2\Delta}(\pi,Y_I) G_n(2\pi (j-j'); -1)\nonumber \\
&=&(2R)^{2\Delta}G_n(2\pi (j-j'))
\eea
where we used the simplified notation 
$G_n(\tau)\equiv G_n(\tau;Y_a \cdot Y_b = - 1)$.
The mutual information in terms of the thermal Green functions is therefore given by:
\be
\label{MI}
I(A,B)=\lim_{n\to 1}\frac{ (4R_AR_B)^{2\Delta }}{2|r_A-r_B|^{4\Delta }}\frac{1}{1-n}\left(\sum_{j=1}^{n-1}G^2_n(2\pi j)\right) + \ldots\,.
\ee
This quantity can be evaluated following similar steps to the calculation of \cite{Faulkner:2014jva} with some obvious modifications.

\begin{figure}[h!]
\centering
\includegraphics[scale=.7]{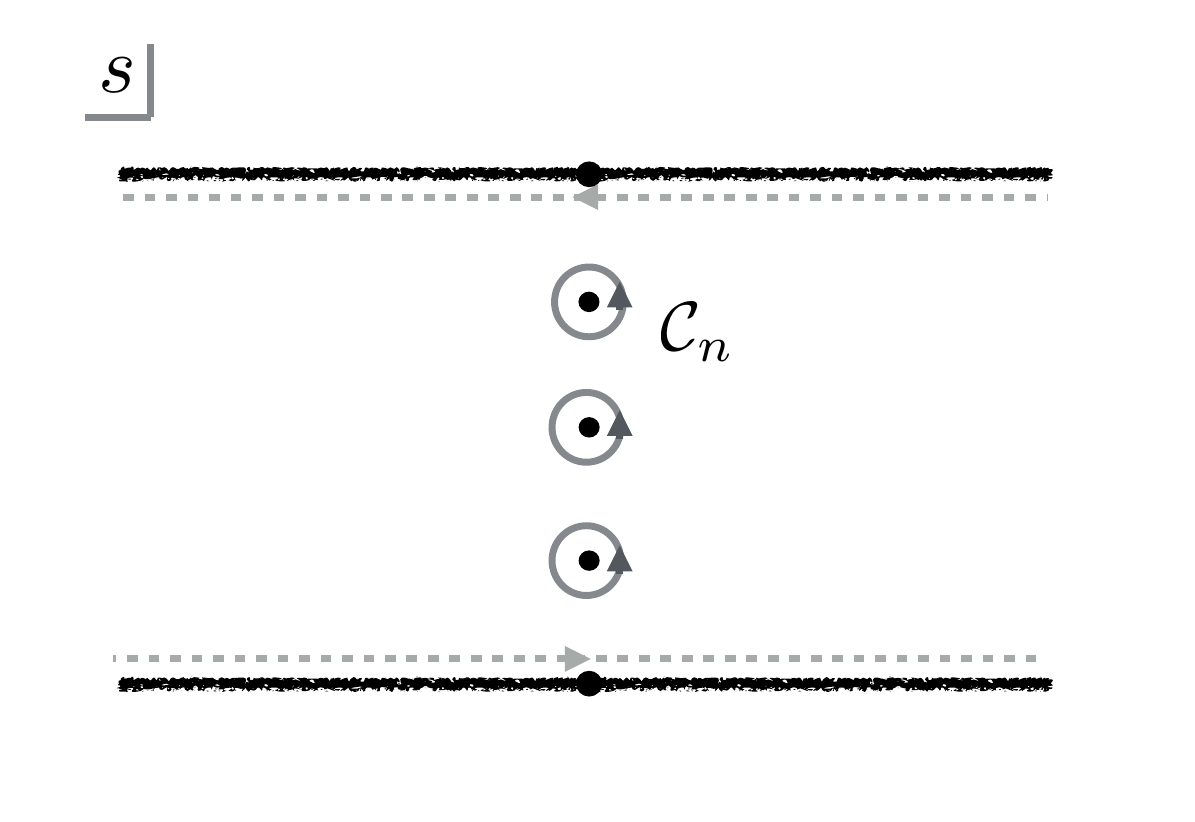}
\caption{Integration contour used to evaluate the sum in \protect\ref{MI}.\label{contour}}
\end{figure}

To start with we use the unique analytic continuation of the thermal green function to the complex time plane $G_n(\tau)\to G_n(-i s)$ with $s\in \mathbb{C}$ and $0<\textrm{Im}(s)<2\pi n$ and expressed the sum as a contour integral 
\bea
\label{analyticconti}
\sigma_n \equiv \sum_{j=1}^{n-1}G^2_n(2\pi j)=\int_{{\cal C}_n} \frac{ds}{2\pi i} \frac{G^2_n(-is)}{e^{s}-1}
\eea
where ${\cal C}_n$ is the prescribed contour in figure \ref{contour}. At this point it is convenient to make a convenient subtraction from the integrand in \eqref{analyticconti} which leads to a vanishing contribution to the integral due to the absence of poles within the contour ${\cal C}_n$:
\be
\sigma_n = \int_{{\cal C}_n} \frac{ds}{2\pi i} G^2_n(-is) k_n(s) \qquad k_n(s) \equiv \left( \frac{1}{e^{s}-1} - \frac{1}{ e^{s/n}-1} \right)
\ee
This subtraction ensures that the kernel $k_n(s)$ vanishes when $n=1$.
Assuming the integrand goes to zero when $\textrm{Re}(s)\to \pm \infty$\footnote{
The assumption boils down to showing that real time thermal correlation function
in Hyperbolic space decays faster than $s^{-1/2}$. Exponential decay is 
natural and this is what one finds at $n=1$ where the correlator decays as $ e^{ - \Delta s}$.  } we can deform the integration contour to the lines with ${\Im } s= \epsilon, 2\pi n - \epsilon$ just above and below the branch cuts that are expected to appear in the thermal Greens function:
\bea
\label{312}
\label{correct}
\sigma_n =\int_{-\infty}^{\infty} \frac{ds}{2\pi i} \left(G^2_n(-is+\epsilon ) k_n(s + i \epsilon) -G^2_n(-is-\epsilon )  k_n( s - i \epsilon) \right)
\eea
where $\epsilon >0$. We have additionally used the $2\pi n$ periodicity of $G^2_n$ and $k_n$ in the complex plane setting $e^{2\pi i n} = 1$ which is true prior to analytic continuation in $n$.\footnote{
The claim is that \eqref{correct} is the correct analytic continuation in $n$ away from the integers.
This is based on the assumption of analyticity in the complex $n$ plane at least for $\Re n >0$
which would not be true had we not dropped the $e^{2\pi i n} = 1$ terms.
See \cite{Faulkner:2014jva} for more discussion on this point.}

The limit of interest can now be taken:
\be
\lim_{n \to 1} \frac{1}{n-1} \sigma_n = \int_{-\infty}^{\infty} \frac{ds}{2\pi i} \left(G^2_1(-is+\epsilon )  \hat{k}(s + i \epsilon) -G^2_1(-is-\epsilon )  \hat{k}( s - i \epsilon) \right)
\ee
where:
\be
\qquad \hat{k}(s) \equiv - \frac{1}{ 4 \sinh^2(s/2)} \, .
\ee
The last step is to deform the integration contour in the first term to ${\Im } s = \pi$ and ${\Im } s = - \pi$
in the second term. This is convenient since $G_1^2( - is' + \pi) = G_1^2(- is' - \pi)$ and these are
non singular when $s'=0$ so we have dropped the $i \epsilon$'s. This leads to the final
answer:
\be
\lim_{n \to 1} \frac{1}{n-1} \sigma_n
 = \int_{-\infty}^\infty d s' \frac{G_1^2( - i s' + \pi)}{ 4 \cosh^2 ( s'/2)} =   \int_{-\infty}^\infty d s' 
 ( 2 \cosh (s'/2))^{-4\Delta -2}
\ee

This last integral is in fact convergent and well defined for $\textrm{Re}(\Delta)>-1/2$ and evaluates to:
\bea
\frac{2\sqrt{\pi}}{4^{2\Delta +1}}\frac{\Gamma(2\Delta +1)}{\Gamma(2\Delta +3/2)}.
\eea
Plugging this result into (\ref{MI}), gives the leading term in the mutual information 
\bea
\label{final3}
I(A,B)=\frac{\sqrt{\pi}\Gamma(2\Delta +1)}{4\Gamma(2\Delta +\frac32)}\frac{ (R_AR_B)^{2\Delta }}{|r_A-r_B|^{4\Delta }} + \ldots
\eea
valid for any CFT. This is a surprising result since it is independent of the space-time dimension of the theory and therefore equals the one for 2D CFTs.

\section{Field Theory in the Bulk \label{4}}

In the previous section we found that the leading term in the mutual information between separated spheres in any dimensions depends only on the lowest scaling dimension of the CFT operators. This powerful result gives us a good amount of data to check the FLM prescription as a reliable method to compute quantum corrections to holographic entanglement entropy. 
This section will focus on carrying out the computation dual to the CFT calculation of the section \ref{3}. That is, we perform the calculation of the leading term in the mutual information between separated hemispheres for a scalar QFT in an AdS background geometry. 

This calculation fits into the framework described in section \ref{2} as applied to an arbitrary QFT in curved geometry. As described there, an important simplification occurs when the theory is a CFT. However, we argue that such a simplification is irrelevant if we are just interested in the leading term of the mutual information at large separation. The relevant coefficient is still given in terms of the two point correlation functions of operators in the conifold AdS background. This quantity can be computed via an application of the method of images, as described in \cite{Cardy:2013nua}, since in this approximation the bulk quantum fields can be considered to be free.

In section \ref{2} the method to extract the appropriate OPE coefficients involved examining multi-point function of a set of fields  in the replicated space at points far from region $A$.
Taking the general result (\ref{correlators}) and applying it to two identical non-unit operators $\phi$ on distinct replicas $j$ and $j'$ this formal expression can be written as 
\bea
\label{41}
\langle \phi^{(j)}(r)\phi^{(j')}(r)\rangle_{{\cal C}^{(n)}_A} 
&=& C_{jj'}^{A} \langle \phi(r)\phi(r_A) \rangle_{{\cal M}}^2 \nonumber \\
&& +\left. \sum_{k_{j}, k_{j'}}\right.' C_{k_j, k_{j'}}^{A}\langle  \phi (r )\Phi_{k_j}(r_A)\rangle_{\cal M}\langle  \phi(r)\Phi_{k_{j'}}(r_A)\rangle_{\mathcal{M}}\,. 
\eea
where this last sum is over operators distinct from $\phi$, that is $\Phi_{k_{j}}, \Phi_{k_{j'}}  \neq \phi$.
Notice that for a CFT the choice of normalization (\ref{normalization}) made the later sum
in \eqref{41} equal to zero unlike in an arbitrary QFT. However, for a free scalar theory, the largest two point function at long distances corresponds to that of the fundamental field $\phi(r)$ with itself, and therefore the two point function between $\phi(r)$ and any other operator of the theory is smaller for large separations. That is, if $r\gg r_A$, then 
\bea
\langle {\Phi}_{k_j, k_j'}(r) \phi (r_A)\rangle \ll \langle \phi(r) \phi (r_A)\rangle\,.
\eea 
That means that the main contribution to  the LHS of (\ref{41}) is given by the first term in the RHS of (\ref{41}). 

 Therefore, by taking $r \to \infty$ we can extract the coefficients $C^A_{jj'}$:
\bea
\label{coeff}
C^A_{jj'}=\lim_{r \to \infty}G_1(r,r_A)^{-1}G_1(r,r_A)^{-1} \langle \phi^{(j)}(r)\phi^{(j')}(r)\rangle_{{\cal C}^{(n)}_A}\,.
\eea
This quantity determines the first correction to the ratio 
\bea
\label{insidelog}
\frac{Z({\cal C}^{(n)}_{A\cup B})Z^n}{Z({\cal C}^{(n)}_{A})Z({\cal C}^{(n)}_{B})}=1+\frac 12\sum_{jj'}{ C}^A_{jj'}{C}^B_{jj'}G_1(r_A,r_B)^{2}+\cdots
\eea
required for the evaluation of the mutual information up to that order. This is the equivalent of (\ref{insidelog1}) in AdS  space-time. 

The two point function for a free scalar in the AdS bulk is given by \cite{D'Hoker:2002aw}
\bea
\label{greenbulk}
G_1(r,r')=\frac{2C_{\Delta}}{\nu}\left(\frac \xi 2\right)^{\Delta}F(\frac \Delta 2, \frac \Delta 2+\frac12;\nu;\xi^2 )
\eea
where 
\bea
\xi=\frac{2zz'}{z^2+z'^2+(t_E - t'_E)^2 + (x-x')^2}\,,
\eea
where the $d$ boundary theory coordinates in euclidean signature are $(t_E,x^\alpha)$
and $x^\alpha$ are the spatial coordinates with $\alpha = 1,\ldots d-1$. 
The metric of AdS is given by:
\be
ds^2 = \frac{ dz^2 + d t_E^2 + d x^2 }{z^2}
\ee
We also have set $\Delta=\frac d2+ \nu$ and $\nu=\sqrt{\frac{d^2}{4}+m^2}$. 

The two points $r_A$ and $r_B$ are well separated such that the Greens function becomes
in this limit
\bea
\label{greenbulkz}
G_1(r_A,r_B)\approx \frac{2C_{\Delta}}{\nu}\frac{z^{\Delta}_A z^{\Delta}_B}{|x_A-x_B|^{2\Delta}} \,.
\eea
We can rewrite (\ref{insidelog}) as
\bea
\label{insidelog2}
\frac{Z({\cal C}^{(n)}_{A\cup B})Z^n}{Z({\cal C}^{(n)}_{A})Z({\cal C}^{(n)}_{B})}=1+\frac 12\sum_{jj'}{ \tilde{C}}^A_{jj'}{\tilde{C}}^B_{jj'}\frac{1}{|x_A-x_B|^{4\Delta}}+\cdots
\eea
where ${\tilde{C}}^A_{jj'}=\frac{2C_{\Delta}z^{2\Delta}_A}{\nu} C^A_{jj'} $ and similarly for $B$.
The mutual information in terms of the tilde coefficients is given by
\bea
\label{mib}
I(A,B)&\approx &\lim_{n\to 1}\frac{1}{1-n}\left(\frac 12 \sum_{j j'} { \tilde{C}^A_{jj'}}{ \tilde{C}^B_{jj'}} \right)\frac{1}{|x_A-x_B|^{4\Delta}}\,. 
\eea

We now  focus on evaluating the coefficients ${\tilde{C}}^A_{jj'}$.
Further simplification is possible in the case in which the entangling surface is a hemisphere in AdS. 
Consider the following inversion transformation
\bea
x'^\alpha&=&\frac{x^\alpha+ n^\alpha R_A}{ (x+R_A n)^2+ t_E^2 + z^2}-\frac{n^\alpha}{2R_A} \,, \qquad
t'_E =\frac{t_E}{(x+R_A)^2+ t_E^2 + z^2}  \,, \nonumber \\ 
z'&=&\frac{z}{(x+R_A n)^2+ t_E^2 + z^2}
\eea
where $n^\alpha = (1,0, \ldots 0)$ is a $d-1$ dimensional vector. 
This transformation maps the conifold ${\cal C}^{(n)}_A$ with singularities located on the hemisphere
$x^2 + z^2 = R_A^2$ at $t_E=0$ to a conifold ${\cal C'}^{(n)}_A$  with singularities located on the
plane $(t'_E = 0, x'^1 = 0)$. This later coordinate system is more suitable for some analytic manipulations.

We
will apply this to the point $r_A$ as well as the point at $r = r_\infty \rightarrow \infty$ needed
to evaluate \eqref{coeff}. The reference point $r_A = (z_A,t_E,x^\alpha)$ we can take in the middle of the $A$ hemisphere:
\be
\left(z = z_A, \, t_E=0, \, x^\alpha = 0 \right) \rightarrow \left( z' = \frac{z_A}{R_A^2 + z_A^2} 
, \, t'_E =0,\, x'^\alpha = \frac{n^\alpha}{2R_A}  \frac{ R_A^2 - z_A^2}{(R_A^2+ z_A^2)}  \right)  
\ee
and the point at infinity becomes:
\be
r_\infty \rightarrow r_\infty' \approx \left( z' =\epsilon 
,  \, t'_E =0, \, x'^\alpha = - n^\alpha/2R_A  \right)  
\ee
where we should take the limit $\epsilon \rightarrow 0$ in order to send $r_\infty \rightarrow \infty$. 
For example in this limit $G_1(r',r'_A) \rightarrow (\epsilon z_A)^\Delta 2 C_\Delta \nu^{-1} $, such that the OPE coefficients in  \eqref{coeff} become 
 \bea
 C^A_{jj'}=\lim_{\epsilon \to 0}\epsilon^{-2\Delta} z_A^{-2\Delta}
 \left(\frac{\nu}{2C_{\Delta}}\right)^2\langle \phi^{({j})}(r')\phi^{(j')}(r')\rangle_{{\cal C'}^{(n)}_A}\,.
\eea 
The tilde coefficients are now
 \bea
 \tilde{C}^A_{jj'}=\lim_{\epsilon \to 0}\epsilon^{-2\Delta}
 \frac{\nu}{2C_{\Delta}}\langle \phi^{(j)}(r')\phi^{(j')}(r')\rangle_{{\cal C'}^{(n)}_A}\,
\eea
Now, we analyticaly continue $n$ to the values $1/m$ where $m$ is an integer. This allows us to calculate the two point function on ${\cal C'}^{(1/m)}_A$ using the method of images since the
resulting space can be regarded as a $\mathbb{Z}_m$ quotient of $AdS$:
\bea
\label{mi}
\langle \phi^{(j)}(r')\phi^{(j')}(r')\rangle_{{\cal C'}^{(1/m)}_A}=\sum_{k=0}^{m-1}\langle \phi(r'[\theta_j + 2 \pi k/m] )\phi(r' [\theta_{j'}] )\rangle_{AdS_{d+1}}\,,
\eea
where we have dropped the $j$ superscript in the fields since now the fields are defined in a single copy of AdS. The point $r'[\theta]$ is defined:
\be
r'[\theta] =\left( z' = \epsilon, \, \, t'_E = - \sin(\theta)/2 R_A , \,\, x'^\alpha = - n^\alpha \cos(\theta)/2 R_A  \right)
\ee
which is a rotation of the point $r'_\infty$ about the conifold plane by angle $\theta$. 
The angle $\theta_j$ will eventually be set equal to $2\pi j$ since they label the replicas when $n$ is  taken to be an integer, however for now we keep $\theta_j$ general.   Using the $z=\epsilon\to 0$ limit of the AdS green functions as well  we can write (\ref{mi}) as 
\bea
\label{mi2}
\langle \phi^{(j)}(r')\phi^{(j')}(r')\rangle_{{\cal C'}^{(1/m)}_A}=\frac{2C_{\Delta}}{\nu}\epsilon^{2\Delta} (2R_A)^{2\Delta}\sum_{k=0}^{m-1}\frac{1}{(2-2\cos(\theta_{j'}-\theta_j+2\pi k/m))^{\Delta}} \,. \nonumber 
\eea
We can calculate this last sum again using  contour integration methods to write 
this as:
\be
\label{away}
g_m(\theta) \equiv \sum_{k=0}^{m-1}\frac{1}{(2-2\cos(\theta+2\pi k/m))^{\Delta}}
= \int_{-\infty}^{\infty} \frac{ds}{2\pi i} \left( q_m(s+ i \delta ) - q_m(s - i \delta) \right)
\ee
where
\be
q_m(s) = \frac{m}{ ( e^{(s - i \theta) m}  -1)} \frac{1}{(2 - 2 \cosh s)^\Delta}
\ee
and $\delta$ is a positive infinitesimal parameter. 
This last expression is then the desired $m$ analytic continuation. Note that
this function is periodic in $\theta \equiv \theta +2\pi/m$ so to make sense
of the continuation in $m$ it is natural to define $\hat{\theta} = m\theta$ which is then
defined for $ 0 < \hat{\theta} < 2\pi$.  In order to gain confidence in this result we study the $n$ continuation more carefully in Appendix A where we check
numerically that the resulting function agrees with the original sum when $m$
is an integer and is well behaved in various limits away from integer $m$. 

We can now take $m \rightarrow 1/n$
where $n$ is integer. We will not need to know explicitly the function $g_{1/n}(\theta)$ just that it satisfies certain nice properties. Firstly it is periodic $g_{1/n}(\theta + 2\pi n)  = g_{1/n}(\theta)$. Secondly when $n=1$ we find:
\be
g_1(\theta) = \frac{1}{ (2 - 2 \cos \theta)^\Delta }
\ee
and finally it is well behaved (decays exponentially) as $s = i \theta \rightarrow \pm \infty$. 
Indeed these properties are sufficient for us
to apply the  same analytic continuation techniques for the replica sum as in Section 3. 
The tilde coefficients become:
\be
\tilde{C}_{jj'} =  (2 R_A)^{2\Delta} g_{1/n} ( 2\pi (j-j') )
\ee
such that the leading correction to the mutual information is:
\be
I(A,B) = \frac{(2 R_A)^{2\Delta} (2 R_B)^{2\Delta} }{2 | x_A - x_B|^{4\Delta}} \lim_{n \to 1} \frac{n}{1-n} \sum_{j=1}^{n-1} 
 (g_{1/n} ( 2\pi j ))^2
\ee
Where this sum is almost identical to \eqref{MI} in section 3. The main difference is that
$G_n$, the CFT greens function on $\mathbb{S}^1\times \mathbb{H}^{d-1}$,  has been 
replaced by $g_{1/n}$. Since $g_{1/n}$ is $2\pi n$ periodic in $\theta$ it can be considered a thermal Green's function just like $G_n$ and thus analytically continued to the complex $\theta$ plane. Repeating the steps of Section 3 and taking the limit $n \to 1$ we arrive at the identical result to the field theory. 
We emphasize that $G_n$ is different from $g_{1/n}$ (the later is known explicitly while the former is highly theory dependent.) The fact that these different thermal green functions $G_n$ and $g_{1/n}$ give rise to the same contribution to the mutual information, is related to the expected agreement between the proposal of \cite{Barrella:2013wja} and the FLM. We have this established this agreement in the situation at hand.

\section{Conclusions \label{5}}
In this work we have presented a set of analytic checks that support the validity of the FLM prescription as applied to the calculation of the leading large distant term in the mutual information between bulk hemispherical regions. This was made possible by providing the CFT counterpart of this calculation, namely, the mutual information between spherical regions in a generic CFT on flat space-time. Notably, the CFT result is universal, and depends only on the lowest scaling dimension of the CFT operators of the theory, and therefore agree with the 2d CFT result. 

The methods used in the calculation of the bulk mutual information are expected to be applicable to the computation of similar contributions 
coming from non-scalar bulk operators, like bulk gravitons. Since there are some critical uncertainties when attempting to apply FLM to gravitons this is an important avenue for further exploration.

\section*{Acknowledgements}
We thank John Cardy, Isaac Cohen-Abbo, Matthew Headrick, Howard Schnitzer, Erik Tonni and Huajia Wang for useful discussions and comments. This work has its roots from conversations with Matthew Headrick and some of the key ideas were also motivated from further interactions with him. Part of this work was done while the authors attended the ``Entanglement in Strongly-Correlated Quantum Matter'' and ``Quantum Gravity Foundations: UV to IR'' workshops at the KITP. We would like to thank the organizers, participants, and KITP staff for a stimulating environment.
C. A. is grateful to Ana Nioradze for encouragement and motivational support during this work. C. A. is supported in part by the DOE by grant DE-SC0009987. C. A is also supported in part by the National Science Foundation via CAREER Grant No. PHY10-53842 awarded to Matthew Headrick.
TF is supported by the DARPA YFA Grant  No. D15AP00108. 

\appendix
   
\section{Analytic continuation of the conifold Green's function}
 
We gave an expression  in \eqref{away}  for the analytic continuation of $g_m(\theta)$ away
from integer $m$ where it can be calculated with the method of images. 
Here we would like to un-package this expression and make some consistency checks.

Firstly the branch cut structure obscures a little the precise definition in \eqref{away} and so
we give here a slightly more refined version by changing integration variables to $\lambda = e^s$ such that:
\be
g_m(\hat{\theta}) =\frac{1}{2\pi i} \int_{C_+ \cup C_-} \frac{d \lambda}{\lambda} \frac{m}{ \lambda^m e^{- i \hat{\theta}} -1}\left[ (-\lambda)^{\Delta} ( 1- \lambda)^{-2\Delta} \right]
\ee 
where we take $C_+ = (0, \infty ) + i\delta$ and $C_- = (\infty,0) - i \delta$ 
and pick the branch cuts of the power functions in the standard way to lie on the negative
real axis. We have set $\hat{\theta} = \theta m$. 
Note that it is \emph{not} convenient to write this as a discontinuity along the branch cut 
for $\lambda >0$ because this discontinuity itself does not integrate to a finite answer (there is a divergence  at $\lambda=1$ for nice values of $\Delta$) and this is regulated by the $i\delta$ prescription. Rather to get a handle on this numerically we  can simply rotate the integration
contour to $\lambda = e^{i \phi} \mu$ where $\phi = {\rm min} [ \hat{\theta}/(2m), \pi ]$
for the $C_+$ contour and $\phi = - {\rm min}[ (2\pi- \hat{\theta})/(2m), \pi ]$ for the $C_-$ contour and we should integrate from $\mu = 0$ to  $\infty$. This choice is to avoid poles in $\lambda$
occurring at the roots: $ \lambda^m e^{ - i \hat{\theta}} - 1 = 0$.  We are assuming $ 0 < \hat{\theta} < 2\pi$ for this discussion which is sufficient because
of the periodicity of $g_m(\hat{\theta})$. 
This is the definition we work with numerically.

We have checked for several values of $\Delta$  that indeed this does agree with the sum in \eqref{away} when $m$ is an integer. It is also clear that $g_m(\hat{\theta})$ is an analytic function of $m$
for ${\rm Re} (m) > 0$ and for large  $m$ behaves as $ \sim m^{2 \Delta}$ which at least gives it some nice properties that one might believe defines it uniquely. 

The final thing to check is the behavior for the real time Greens function: $\hat{\theta} \rightarrow i \eta$ and taking $ \eta \rightarrow \pm \infty$. It is not hard to see that:
\be
 \lim_{\eta \rightarrow \pm \infty} g_m(i \eta) \propto e^{ - |\eta| \Delta/m} 
\ee
All of these properties makes this a well defined euclidean thermal green function that
we could for example compare to $G_{1/m}$ defined in Section 3.

\newpage
\bibliographystyle{utphys}

\bibliography{emirefs}

\end{document}